# Acoustic square-root topological insulators

Mou Yan[1*], Xueqin Huang[1*], Li Luo[1], Jiuyang Lu[1†], Weiyin Deng[1†], and Zhengyou Liu[2,3†]

[1]School of Physics and Optoelectronics, South China University of Technology, Guangzhou 510640, China
[2]Key Laboratory of Artificial Micro- and Nanostructures of Ministry of Education and School of Physics and Technology, Wuhan University, Wuhan 430072, China
[3]Institute for Advanced Studies, Wuhan University, Wuhan 430072, China

*M.Y. and X.H. contributed equally to this work
†Corresponding author. Email: phjylu@scut.edu.cn; dengwy@scut.edu.cn; zyliu@whu.edu.cn

Square-root topological states are new topological phases, whose topological property is inherited from the square of the Hamiltonian. We realize the first-order and second-order square-root topological insulators in phononic crystals, by putting additional cavities on connecting tubes in the acoustic Su-Schrieffer-Heeger model and the honeycomb lattice, respectively. Because of the square-root procedure, the bulk gap of the squared Hamiltonian is doubled. In both two bulk gaps, the square-root topological insulators possess multiple localized modes, i.e., the end and corner states, which are evidently confirmed by our calculations and experimental observations. We further propose a second-order square-root topological semimetal by stacking the decorated honeycomb lattice to three dimensions.

In general, topological insulators (TIs) are characterized by a bulk gap and in-gap boundary states [1]. The conventional $d$-dimensional ($d$D) TIs are featured with $(d-1)$D boundary states [2,3]. For instance, the 1D TIs, such as the Su-Schrieffer-Heeger (SSH) model, host 0D end states; the 2D TIs host 1D edge states, such as the quantum spin Hall insulators. Recently the higher-order TIs exhibit an extended bulk-boundary correspondence, namely, a $d$D $n^{\text{th}}$-order TIs have $(d-n)$D boundary states [4-6]. For example, a 2D second-order TI possesses the 0D corner states. These intriguing topological physical effects have been extended to artificial band gap materials including photonic crystals [7-20] and phononic crystals (PCs) [21-32]. Due to the high-precision of sample fabrication, the classical wave systems provide an eminent platform to explore intriguing topological properties, which in turn provide novel manipulation for light and sound waves, such as the topological insulator laser [33,34] and topological negative refraction [35]. To date, searching and realizing new topological states in different systems continue to attract a lot of attention.

Recently, a square-root TI (SRTI) is proposed with its topological property inherited from the parent Hamiltonian [36-39]. The square-root procedure had played a seminal role in deriving the Dirac equation for electron in relativistic quantum mechanics from the quadratic Klein-Gordon equation, which revealed the built-in chirality for electron. This procedure provides an antidiagonal block matrix form and leads to a symmetric spectra for square-root Hamiltonian. During the procedure, the square-root Hamiltonian will inherit the topological property of the parent Hamiltonian, thus give rise to the concept of SRTI [36]. For lattice models, SRTIs, including the first-order [37] and higher-order [38], can be generated by inserting additional sites and separating all the original ones. So far, only a first-order SRTI has been realized very recently, in a photonic chain of Aharonov-Bohm cages with magnetic flux [39].

In this work, we demonstrate the first-order and second-order SRTIs in PCs, by inserting a set of additional cavities in the acoustic SSH model and decorated honeycomb lattice, respectively. We found 0D end modes for the 1D SRTI and 0D corner modes for the 2D SRTI. Thus, the two PCs are respectively the first-order and second-order acoustic TIs from square roots. In terms of SRTI, the topological property for the 1D PC is inherited from the parent acoustic SSH model, while for the 2D PC, the topological property is induced from the square-root procedure and inherited from inserted lattice, which further extends the scheme in the construction of a square-root

Hamiltonian. Throughout this work, all simulations are performed by COMSOL Multiphysics, and the calculated results are well consistent with the experimental measurements.

We start to construct the SRTI chain from the 1D SSH model, which is shown in Fig. 1(a). The SSH chain consists of two nonequivalent sites in each periodic cell. The intercell and intracell coupling strengths are $t_0(1-\eta)$ and $t_0(1+\eta)$, where $\eta$ ($|\eta| \leq 1$) denotes the deviation from unity and hereafter $t_0 = 1$ without loss of generality. Shown in the lower panel of Fig. 1(a), the SRTI chain is designed by inserting another two sides, which separate the original ones. The new lattice model is right the square root of the SSH model, when the coupling strengths take the square-root values, i.e., $t_1 = \sqrt{1-\eta}$ and $t_2 = \sqrt{1+\eta}$. This is illustrated by squaring the Hamiltonian of the SRTI model,

$$H_{1D}^2 = \begin{pmatrix} 0 & h_{1D} \\ h_{1D}^\dagger & 0 \end{pmatrix}^2 = 2I_{4\times 4} + 2h_{SSH} \oplus h_R, \tag{1}$$

where the antidiagonal block in the square-root Hamiltonian $H_{1D}$ is

$$h_{1D} = \begin{pmatrix} e^{ik/4}\sqrt{1-\eta} & e^{-ik/4}\sqrt{1+\eta} \\ e^{-ik/4}\sqrt{1+\eta} & e^{ik/4}\sqrt{1-\eta} \end{pmatrix}$$

with its complex-conjugate denoted as $h_{1D}^\dagger$. In Eq. (1), the square of the $H_{1D}$ equals the direct sum of the Hamiltonians of a SSH model, $h_{SSH} = \cos(k/2)\sigma_x + \eta \sin(k/2)\sigma_y$, and a residual Hamiltonian, $h_R = \sqrt{1-\eta^2}\cos(k/2)\sigma_x - \eta\sigma_z$, apart from an identity matrix of order four. The identity term globally shifts the eigen values and ensures the positive-semidefinite squared Hamiltonian. The $h_{SSH}$, as the parent Hamiltonian of $H_{1D}$, attribute to the nontrivial topological property when $\eta > 0$, i.e., the coupling strength of intracell larger than that of the intercell.

A finite chain with its bulk Hamiltonian described $H_{1D}$ then hosts end states for $\eta > 0$. We focus on the case $\eta = 1$, i.e., $t_1 = 0$ and $t_2 = \sqrt{2}$, where the outmost two sites (coupled by $t_2$) are detached completely from the bulk. Thus, the chain is trimerized in its bulk and dimerized at the ends, leading to the non-trivial topological phase described by the polarized Wannier center as that for the SSH chain, which is represented by topological end states appearing at the energies $\pm\sqrt{2}$, as shown in Fig. 1(b). Because the energy spectrum is adiabatically connected at each $\eta > 0$, the end states are topological boundary states.

For the 2D lattice, we design a SRTI by decorating the honeycomb lattice as shown in Fig. 1(c). The coupling between the inserted sites are chosen to be the same as above $t_1$ and $t_2$. For simplicity, we introduce the real-space vectors $\boldsymbol{\delta}_1 = (0, \sqrt{3}/6)$ and $\boldsymbol{\delta}_{2,3} = (\mp 1/4, -\sqrt{3}/12)$, pointing from one site of the honeycomb lattice to three nearest inserted sites. Because of the decorating sites and the original are full separated, the $5 \times 5$ Hamiltonian $H_{2D}$ of this lattice possess an antidiagonal block matrix form, akin to the 1D case. Therefore, the square-root feature of $H_{2D}$ can be demonstrated as

$$H_{2D}^2 = \begin{pmatrix} 0 & h_{2D} \\ h_{2D}^\dagger & 0 \end{pmatrix}^2 = h_{2D}h_{2D}^\dagger \oplus h_{2D}^\dagger h_{2D}, \qquad (2)$$

where

$$h_{2D} = \begin{pmatrix} e^{i\boldsymbol{k}\cdot\boldsymbol{\delta}_1}\sqrt{1-\eta} & e^{i\boldsymbol{k}\cdot\boldsymbol{\delta}_2}\sqrt{1-\eta} & e^{i\boldsymbol{k}\cdot\boldsymbol{\delta}_3}\sqrt{1-\eta} \\ e^{-i\boldsymbol{k}\cdot\boldsymbol{\delta}_1}\sqrt{1+\eta} & e^{-i\boldsymbol{k}\cdot\boldsymbol{\delta}_2}\sqrt{1+\eta} & e^{-i\boldsymbol{k}\cdot\boldsymbol{\delta}_3}\sqrt{1+\eta} \end{pmatrix}.$$

In Eq. (2), $h_{2D}h_{2D}^\dagger$ is a $2 \times 2$ matrix, which equals, apart from an identity matrix, to the Hamiltonian $h_H$ of the parent honeycomb lattice with different onsite potentials $\mp 3\eta$ on two sublattices, i.e., $h_{2D}h_{2D}^\dagger = 3I_{2\times 2} + h_H$, while $h_{2D}^\dagger h_{2D}$ is a $3 \times 3$ matrix with the non-trivial part equals to the Hamiltonian $h_K$ of an breathing Kagome lattice, i.e., $h_{2D}^\dagger h_{2D} = 2I_{3\times 3} + h_K$. Remarkably here, the topological property of the decorated honeycomb lattice is not inherited from its parent honeycomb lattice, but induced from the inserted breathing Kagome lattice. Specifically, when $\eta > 0$, the breathing Kagome lattice is of a non-trivial topological phase, which hosts corner states of zero energy originating from the mismatch of the Wannier center to the lattice site. Accordingly, the decorated honeycomb lattice is topologically non-trivial, and further because it is a SRTI the corner states appear with non-zero energies $\pm\sqrt{2}$ show in Fig. 1(d). Note that any one of three edges for the finite 2D lattice in Fig. 1(c) can be degraded to the 1D SRTI chain when the coupling $t_1 = \sqrt{1-\eta}$ vanishes, which is consistent with the extreme condition $\eta = 1$ for a nontrivial phase. The adiabatically connected energy spectrum thus leads to the topological corner states with the energies robust to the coupling parameter $\eta$, as in the above 1D case.

We now consider the acoustic realizations of these SRTIs. Figure 2(a) shows the 1D acoustic SRTI fabricated through 3D printing. Acoustic cavities and tubes have been mapped as the sites and couplings in the lattice model. Here, to construct the acoustic SRTI, additional cavities are inserted in the original acoustic SSH chain. Since we are considering the dipole cavity mode, each pair of nearest-neighbor cavities are connected by two identical tubes. The coupling strength can be tuned by varying the diameter of the connecting tubes indicated as $d_1$ and $d_2$ in Fig. 2(a). The topological properties are different in our acoustic system for the two cases of $d_1 > d_2$ and $d_1 < d_2$. Specifically, we choose $d_1 = 0.8$ cm, $d_2 = 1.2$ cm that corresponding to the topological nontrivial phases.

To demonstrate the end states of the acoustic SRTI, we numerically calculated eigen frequencies and eigen modes with 20 unit cells in Fig. 2(b). As expected, two pairs of degenerate end states are found within the bulk band gaps at $f = 4.20$ kHz and $f = 5.27$ kHz. The two insets are the eigen modes of the end states at the two different frequencies that the acoustic pressures are highly (in-phase and anti-phase) localized at the rightmost two cavities. The other two end states are localized at the left end. When switching to $d_1 = 1.2$ cm and $d_2 = 0.8$ cm, the system transits to trivial phase and no end states exist. In experiment, we obtain the distribution of acoustic pressure in the finite structure (7 periods) by exciting and detecting at each cavity. Figure 2(c) shows the normalized pressure distributions at the frequency of 4.70 kHz and 5.27 kHz, corresponding to the bulk states (left panel) and end states (right panel). For end states, the measured acoustic pressures at ends are much larger than those at the bulk lattices, which is consistent with the simulation results.

A 2D acoustic higher-order SRTI can be engineered by inserting additional cavities in the acoustic honeycomb lattice that is shown in Fig. 3(a). The sizes of inserted cavities keep the same to original ones. The coupling tubes with diameters $d_1$ and $d_2$ are designed different as shown in the inset, thus breaking the lattice symmetry $C_{3v}$ and opening the band gaps. Based on the decorated honeycomb lattice, we construct a triangular acoustic structure containing 81 cavities in total and numerically calculate the eigen frequencies and eigen modes in Fig. 3(b). In the upper and lower band gaps, there are two groups of triply degenerate corner states (red dots) at the frequencies 4.05 kHz and 5.09 kHz, which is different from the well-known Kagome lattice that the corner states are pined to the zero energy. These degenerate corner states are localized on the three corners of the triangular structure. Similar to the end states of 1D SRTI, the two groups of corner states are localized at the outmost two cavities, and the acoustic pressure distribution of eigen modes are featured by the in-phase and anti-phase distributions in the two cavities. Near the bulk states (gray dots), the edge states (blue dots) are localized on the edge cavities of the triangular sample.

To measure the bulk, edge and corner states, we put the exciting source in a cavity marked by the yellow star in Fig. 3(a), and obtain the acoustic pressure in cavities B, E and C, that are away from the same distance from the source. The transmission spectra for B (gray), E (blue) and C (red) are acquired in Fig. 3(c), it is obvious that four peaks appear in the bulk band gaps, corresponding to the corner states at frequencies 4.09 kHz and 5.12 kHz, and edge states around 3.97 kHz and 5.30 kHz. The tiny frequency shifts between the experiment and simulation ($< 1\%$) mainly from the deviation of structure fabrication. In addition, we change measurement method by moving the source and probing the acoustic pressure in each cavity as mentioned above, and then map the distribution of acoustic pressure at the frequencies of corner, edge and bulk states in Figs. 3(d)-3(f). These results match well with the simulation and undoubtedly demonstrate the acoustic second-order SRTI.

Furthermore, we can construct a second-order square-root topological semimetal by stacking the decorated honeycomb lattice layer by layer along the $z$ direction, which is similar to the construction of the second-order topological semimetal from the 2D second-order TI [5,40]. The interlayer coupling $t_z$ is introduced between the one of the honeycomb sites and the inserted sites, which in effect makes a replacement, such as $t_1 \to t_1 + 2t_z \cos k_z$ in the Hamiltonian $H_{2D}$. The non-zero energy band gaps close at $t_1 + 2t_z = t_2$ forming the critical Weyl points with one zero-slope dispersion along the $k_z$ direction. The square-root property is inherited by this semimetal with $t_1 + 2t_z > t_2$ and the topological corner states appear for $k_z$ satisfying $t_1 + 2t_z \cos k_z < t_2$. This lattice model can be implemented in PC as the layer stacking of acoustic second-order SRTI, which may stimulate the further explorations of the square-root systems.

In conclusion, we have realized the first-order and second-order SRTIs. The end and corner modes in two bulk gaps, inherited from the squared first-order and second-order TIs, are evidently observed in experiment. Note that our PC samples are ideal to exhibit the square-root features in contrast to the electronic systems, where simultaneous measurement of the boundary states in two bulk gaps is difficult due to the fermionic band-filling. Our work extends the topological materials to the square-root ones, thus is of significant interest and may offer possible applications for advanced acoustic devices such as sensing or acoustic manipulation.

**Acknowledgements**

This work is supported by the National Natural Science Foundation of China (Nos. 11890701, 11804101, 11704128, 11774275, 11974120, 11974005), the National Key R&D Program of China (No. 2018YFA0305800), the Guangdong Basic and Applied Basic Research Foundation (No. 2019B151502012), the Guangdong Innovative and Entrepreneurial Research Team Program (No. 2016ZT06C594), the National Postdoctoral Program for Innovative Talents (BX20190122), the China Postdoctoral Science Foundation (2019M662885).

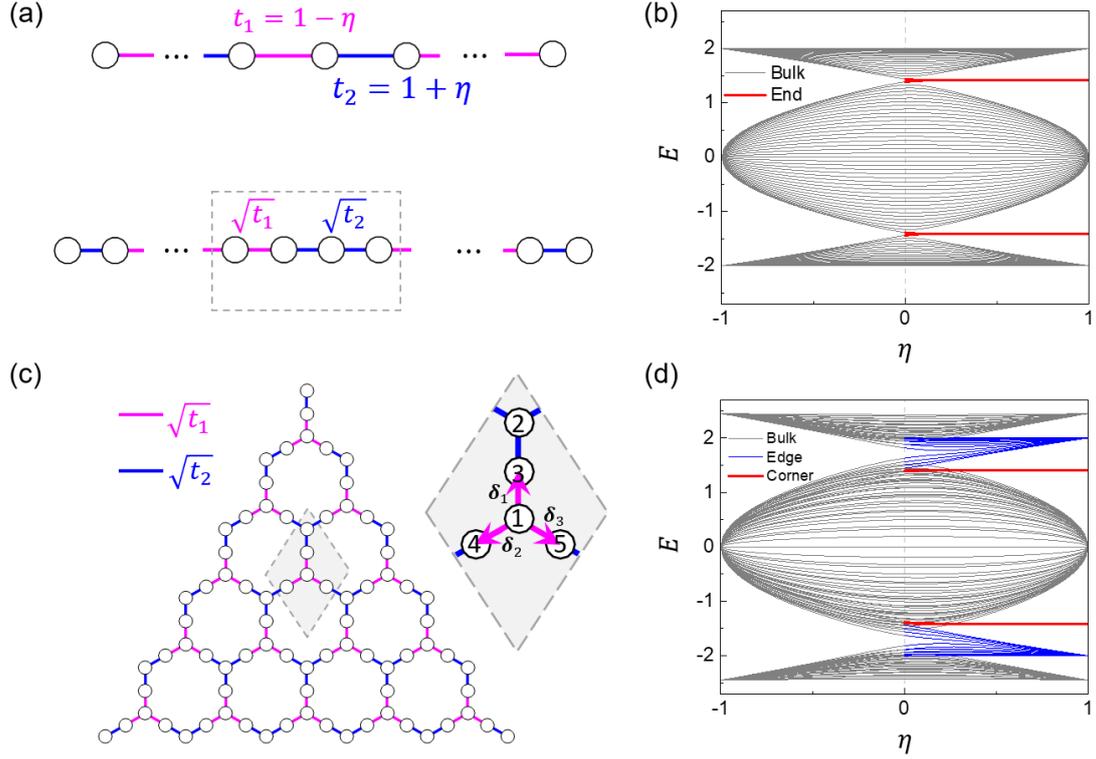

FIG. 1. Lattice models for SRTIs. (a) Top panel: SSH model with $t_1$ and $t_2$ denoting the intracell (blue) and intercell (pink) couplings. Bottom panel: the 1D first-order SRTI by inserting an additional site on each connection (black dot). The couplings of the additional site and its nearest neighbor ones are $\sqrt{t_1} = 1-\eta$ and $\sqrt{t_2} = 1+\eta$ respectively. The gray dashed boxes mark one unit cell. (b) Energy spectrum of a finite 1D SRTI as a function of $\eta$. The gray and red lines represent the bulk and end states, respectively. (c) Lattice model of the 2D second-order SRTI constructed from the honeycomb lattice with additional sites on each connection. (d) Energy spectrum as a function of $\eta$, the gray, blue and red lines denote the bulk, edge and corner states, respectively.

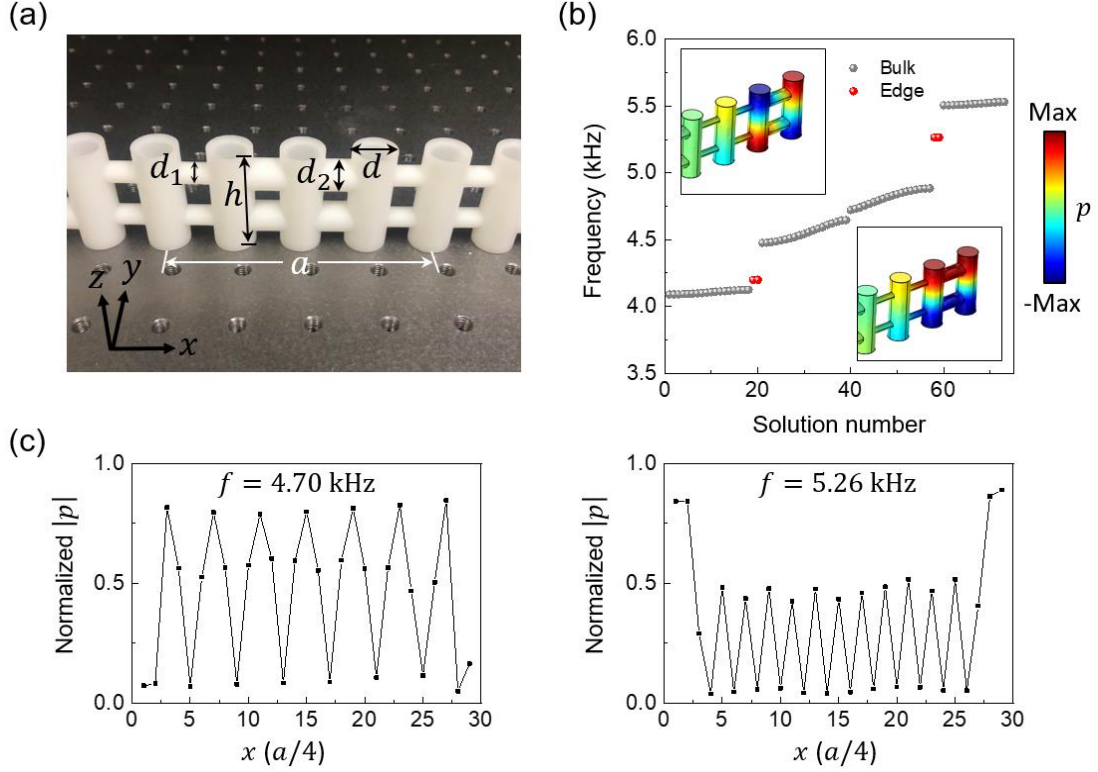

FIG. 2. Realization of the 1D first-order SRTI in PC. (a) Photo of acoustic SRTI sample consisting of connected identical cylindrical cavities. The lattice constant is $a = 10$ cm, the height and diameter of each cavity are $h = 4.2$ cm and $d = 1.6$ cm. The diameters of the connections of the cavities are $d_1 = 0.8$ cm and $d_2 = 1.2$ cm. (b) Numerically simulated eigen frequencies of a finite sample composed of 20 unit cells. The gray and red spheres represent the bulk and end states, respectively. Top (Bottom) inset: the eigen pressure distribution for the right (left) end state at a frequency of 4.20 kHz (5.27 kHz). (c) Measured pressure distributions of the bulk (left panel) and end (right panel) states.

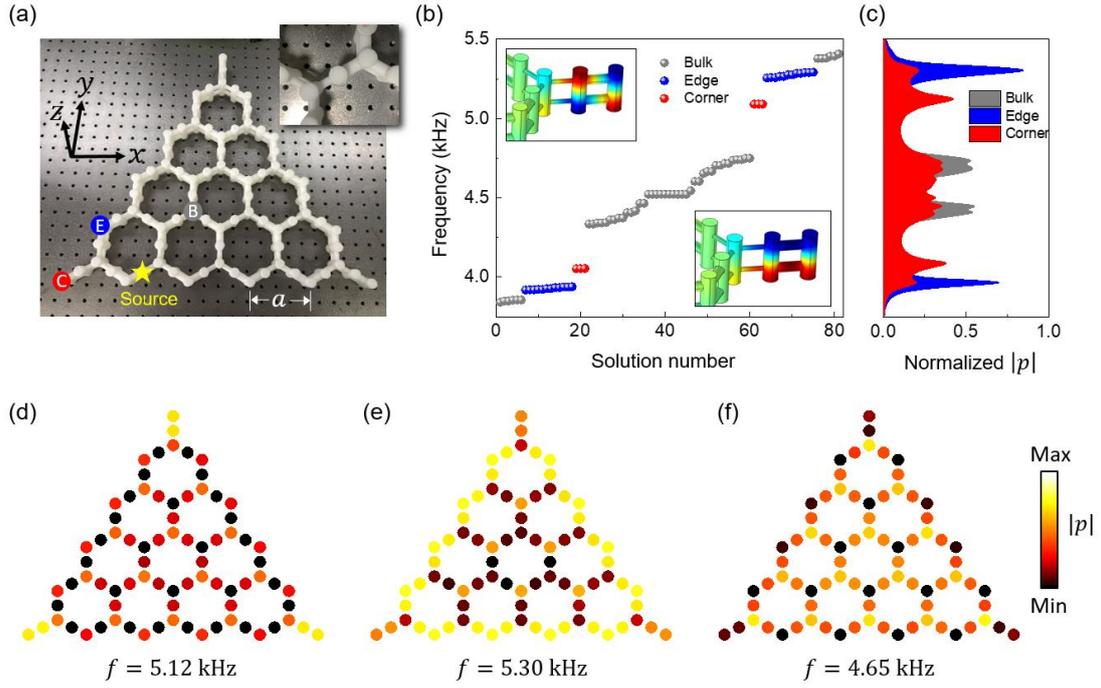

FIG. 3. Realization of the 2D second-order SRTI in PC. (a) Photo of the sample. Inset: the enlarged view of the unit cell. (b) Numerically simulated eigen frequencies of the finite second-order SRTI. The gray, blue and red spheres represent the bulk, edge and corner states, respectively. Top (Bottom) inset: the eigen pressure distribution of the corner state at the frequency of 5.09 kHz (4.05 kHz). (c) The measured response spectrums of the bulk (gray), edge (blue) and corner (red) states, corresponding to the detecting points B, E and C. The yellow star marks the position of the source. (d-f) The measured pressure field distributions of the corner, edge and bulk states.